# Improved Algorithm for Degree Bounded Survivable Network Design Problem

Anand Louis[*]        Nisheeth K. Vishnoi[†]

October 22, 2018

**Abstract**

We consider the *Degree-Bounded Survivable Network Design Problem*: the objective is to find a minimum cost subgraph satisfying the given connectivity requirements as well as the degree bounds on the vertices. If we denote the upper bound on the degree of a vertex $v$ by $b(v)$, then we present an algorithm that finds a solution whose cost is at most twice the cost of the optimal solution while the degree of a degree constrained vertex $v$ is at most $2b(v) + 2$. This improves upon the results of Lau and Singh [13] and Lau, Naor, Salavatipour and Singh [12].

---

[*]College of Computing, Georgia Tech., Atlanta. This work was done while the author was affiliated with Indian Institute of Technology Delhi, New Delhi, and was visiting Microsoft Research India, Bangalore. Email: `anand.louis@cc.gatech.edu`

[†]Microsoft Research India, Bangalore. Email: `nisheeth.vishnoi@gmail.com`

# 1 Introduction

**The degree-bounded survivable network design problem.** In the Survivable Network Design Problem (SNDP), the input is an undirected graph $G = (V, E)$ and connectivity requirements $\rho(u, v)$ for all pairs of vertices $u, v$. The goal is to select a subset of edges from $E$ such that there are $\rho(u, v)$ edge disjoint paths between every $u, v \in V$. In addition, if each edge has an associated cost, then the goal is to find the minimum cost solution satisfying the connectivity requirements. Interest in this problem derives from its applications in algorithm design, networking, graph theory and operations research. This problem is NP-hard and the best approximation algorithm is, a 2-approximation, due to Jain [8]. In this paper we will consider the degree bounded variant of the SNDP. Here, in addition we are given degree constraints $b : W_0 \mapsto \mathbb{Z}_{\geq 0}$ on a subset of vertices $W_0$. The goal is to find a network of minimum cost satisfying the connectivity requirements and, in addition, ensuring that every vertex in $W_0$ has degree at most $b(v)$. This problem is referred to as Minimum Degree-Bounded Survivable Network Design Problem (Deg-SNDP).

**The iterative rounding approach.** Jain's approach for SNDP starts off by writing a natural linear programming relaxation for it, with a variable for each edge, and considering an optimal solution to it. In fact, the key to his approach is to consider a *vertex optimal solution* to his LP and derive enough structural information from this solution to prove that it has a variable of value at least $1/2$.[1] Then, picking this edge and noticing that the residual problem is also an instance of a (slight generalization) of SNDP, one could iterate and obtain a solution with cost at most 2 times the cost of the optimal solution. Fleischer *et. al* [4] have generalized this to the element connectivity SNDP problem.

For Deg-SNDP, one can easily augment Jain's LP with degree constraints on the vertices. It is no longer clear that a vertex optimal solution to this LP should have a variable of value at least $1/2$. It was proved by Lau, Naor, Salavatipour and Singh [12] that in any vertex optimal solution to this degree-constrained LP, either there is an edge whose variable has value at least $1/2$ or there is a degree constraint vertex such that at most 4 edges are incident to it in this solution. Using this iteratively[2], they obtain an algorithm which outputs a solution of cost at most 2 times the optimal and such that the degree of every vertex in this solution is at most $2b(v) + 3$.

**Necessity of degree bound violations?** Improving the 2 in the approximation factor would result in improving the approximation for the SNDP problem itself and it seems out of bound. We study to what extent can the $2b(v) + 3$ be improved. As shown by Lau *et al.* [12], the issue with degree constraints here is that there are instances such that this LP may be feasible even when there may be no feasible integral solution to the problem: consider a 3-regular, 3-edge-connected graph without a Hamiltonian Path. Let $W_0 = V$, $b(v) = 1$ for all $v \in V$ and $\rho(u, v) = 1$ for all $u \neq v \in V$. It can be seen that $x_e = 1/3$ is a feasible solution to this LP. On the other hand, it is easy to observe that there is no integral solution even when one relaxes the degree bound function from $b(v) = 1$ to $2b(v) = 2$ for every $v \in V$. This is because such a feasible integral solution would correspond to a Hamiltonian Path. Hence, in any feasible integral solution, there must be a vertex $v$ of degree at least $2b(v) + 1$ (or $b(v) + 2$). Further, Lau and Singh [13] show a family of instances which have a feasible fractional solution but every integral solution to an instance from their family has a degree constrained vertex $v$ with degree $b(v) + \Omega(\max_{u,v \in V} \rho(u, v))$. Hence, the best we can hope for, in general, is an approximation algorithm based on this LP which outputs a solution of cost at most 2 times that of the optimal of this LP and though it satisfies all the connectivity requirements, it satisfies the relaxed degree constraints: for every vertex, the degree in solution obtained is at most $2b(v) + 1$.

---

[1] This is not guaranteed at a non-vertex optimal solution.

[2] Since intermediate instances arising in their algorithms could have semi-integral degree bounds, they need to consider a more general input where the degree bounds could be half-integral.

**Our result.** We improve the result of [12, 13] for Deg-SNDP to $2b(v) + 2$. More precisely, we present an algorithm that finds a solution whose cost is at most twice the cost of the optimal solution while the degree of a degree constrained vertex $v$ is at most $2b(v) + 2$.

An important special case of the Deg-SNDP problem is the Minimum Bounded Degree Spanning Tree Problem. The breakthrough work of Goemans [6] was followed up by Singh and Lau [17] to provide an optimal, in terms of the degree violations, LP-based result for this problem. For Deg-SNDP, proving the exact limits of the LP-based approach has been significantly more challenging as is evident from the work of [12, 13]. This paper leaves us just one step away from this goal.

**Overview and technique.** We follow the iterative rounding approach of Jain. We start with the LP augmented with degree constraints on a set of vertices $W_0$ given as input. Initially $b(v)$ is an integer for all $v \in W_0$. Since in an intermediate iteration we may be forced to pick an edge of value $1/2$, we will have to work with the degree constraint function $b$ which is half-integral. The instance at any intermediate iteration consists of the edges which have neither been *picked* in or *dropped* from any previous iteration along with the suitably modified connectivity requirement function and, the set of vertices with degree constraints is a subset of $W_0$. The latter are vertices from which the degree constraint has not been removed in any previous iteration.

As is usual, during any iteration, we first find a vertex optimal solution $x^\star$ to the current instance. One way to prove our result would have been to tighten the Lau *et al.* result to show that in any intermediate iteration we can either find an edge $e$ such that $x_e^\star \geq 1/2$ or that there is some degree constrained vertex with at most 3 edges in $x^\star$ with value strictly bigger than 0. Unfortunately, we do not know how to prove that. Instead, we are able to prove the following which suffices to prove our result. Let $E_{>0}$ denote those edges with $x_e^\star > 0$ and let $W$ denote the current set of degree constrained vertices and $b$ be the degree constraint function on them. Then, one of the following holds:

- there is an edge $e$ such that $x_e^\star = 0$,
- or there is an edge $e$ such that $x_e^\star = 1$,
- or there is an edge $e = \{u, v\}$ such that $1/2 \leq x_e^\star < 1$, and if $u \in W$, then $b(u) > 1$, and if $v \in W$, then $b(v) > 1$,
- or there is a vertex $v \in W$ such that $\deg_{E_{>0}}(v) \leq 2b(v) + 2$.

When we find an edge of value at least $1/2$ but strictly less than 1, and we decide to pick it, we need to reduce the degree of its endpoint(s) if they are constrained. To maintain feasibility, one can only reduce these degree constraints by at most the value of the edge picked. Since, we know that this is at least $1/2$, we instead choose to reduce the degree constraint by exactly $1/2$. Our invariant allows us to make sure that for any such edge we find, its endpoints, if degree constrained, never drop below 1. Hence, $b(v)$ never becomes less than 1. This is not guaranteed by the invariant of Lau *et al.* Moreover, we remove the degree constraint from a degree constrained vertex $v$ when $\deg_{E_{>0}}(v)$ falls below $2b(v) + 2$. The invariant along with this *iterative relaxation* of constraints are crucial to the analysis of the algorithm. The main task then becomes to prove this invariant. This is done by a new *token distribution* argument the details of which, due to their highly technical nature, are left to a later section.

**Other related work.** Network design problems with degree constraints have been extensively studied by researchers in the computer science and operations research community. A well known problem is the Minimum Bounded Degree Spanning Tree (MBDST) problem where the objective is to minimize the cost of the spanning tree while satisfying the degree constraints on the vertices. This problem is NP-hard as solving an instance of this problem having degree bound of 2 for each vertex would be equivalent to solving the Hamilton Path problem on that graph. This problem had been studied extensively in [5, 6, 17, 3]. The best known algorithm is due to Singh and Lau [17] which returns a solution of cost atmost the cost of OPT and degree of a degree bounded vertex is at most 1 more than its degree bound (see [17]).

The technique of iterative rounding was first introduced by Jain [8] and has been used subsequently in deriving several other important results such as [9, 12, 13, 1, 15]. The special case of SNDP with metric costs has been studied in [7, 2]. Other related network design problems have been studied in [10, 11, 5]. For a detailed review of iterative rounding, the reader is referred to [14, 16].

**Organization.** We start with formal definitions, the LP relaxation for Deg-SNDP in Section 2. We present our main theorem and our algorithm in Section 3. In this section we prove how the main theorem is implied by our main lemma about the vertex optimal solution. Section 4 is devoted to the proof of the main lemma.

## 2 Preliminaries

Given an undirected graph $G = (V, E)$ with $|V| = n$, a subset of vertices $W_0$, a cost function on the edges $c \colon E \to \mathbb{R}_{\geq 0}$, a *connectivity requirement* function $\rho \colon V \times V \to \mathbb{Z}_{\geq 0}$, and a *degree bound* function $b \colon W_0 \to \mathbb{Z}_{\geq 0}$, the goal in the Minimum Bounded Degree Survival Network Design Problem (Deg-SNDP) is to find a subset of edges $H$ of $G$ of minimum cost such that, for every $u, v \in V$, there are $\rho(u, v)$ edge-disjoint paths connecting $u$ and $v$ in the graph on edges in $H$ and the degree of every $v \in W_0$ in $H$ is at most $b(v)$. Here, $\mathsf{cost}(H) \stackrel{\text{def}}{=} \sum_{e \in H} c(e)$. For a set $S \subseteq V$, let $R(S) \stackrel{\text{def}}{=} \max_{u \in S, v \in \bar{S}} \rho(u, v)$. The function $R$ is *weakly super-modular*[3], and in general we will assume the weaker property that the function $R$ is weakly super-modular, rather than derived from some connectivity function $\rho$. Hence, we will often denote an instance of Deg-SNDP by a tuple by $(G(V, E), W_0, c, R, b)$ where $R$ is any weakly super-modular function from $2^{[V]}$ to $\mathbb{Z}_{\geq 0}$.

For an instance $\mathcal{I} = (G(V, E), W_0, c, R, b)$ of Deg-SNDP, let $\mathsf{opt}(\mathcal{I})$ denote the cost of the optimal solution. Let $\mathsf{lp}(\mathcal{I})$ denote the value of the LP of Figure 1 for $\mathcal{I}$.

$$
\begin{aligned}
\mathsf{lp}(\mathcal{I}) \stackrel{\text{def}}{=} \quad \text{minimize} \quad & \sum_{e \in E} c(e) x_e \\
\text{subject to} \quad & \\
\forall S \subseteq V, \quad & x(\delta_E(S)) \geq R(S) \\
\forall v \in W_0, \quad & x(\delta_E(v)) \leq b(v) \\
\forall e \in E, \quad & x_e \geq 0
\end{aligned}
$$

Figure 1: The LP for Deg-SNDP Problem

---

[3] A function $f \colon 2^{[V]} \mapsto \mathbb{Z}_{\geq 0}$ is said to be *weakly super-modular* if $f(V) = 0$, and for every two sets $S, T \subseteq V$, at least one of the following conditions holds:

- $f(S) + f(T) \leq f(S \setminus T) + f(T \setminus S)$
- $f(S) + f(T) \leq f(S \cap T) + f(S \cup T)$.

Here $x(\delta(S)) \stackrel{\text{def}}{=} \sum_{e \in \delta_E(S)} x_e$. Where, for a set $S \subset V$, and a collection of edges $F$ on $V$, $\delta_F(S)$ denotes the subset of edges of $F$ with one endpoint in $S$ and the other in its complement. Also, for a set $S \subseteq V$, let $\chi_S \in \{0, 1\}^E$ be the vector such that $\chi_S(e = \{i, j\}) = 1$ if $|\{i, j\} \cap S| = 1$ and $\chi_S(e) = 0$ otherwise. It is easily seen that this LP is a relaxation to Deg-SNDP. Also, it is known (see [8]) that this LP has a polynomial time separation oracle.

We call an algorithm a $(\alpha, \beta, \gamma)$-approximation for Deg-SNDP if on every instance $\mathcal{I}$ of it, the algorithm outputs a collection of edges which have cost at most $\alpha \cdot \text{opt}(\mathcal{I})$, satisfy the $R$ constraints, and the degree of every vertex in $v \in W$ is at most $\beta \cdot b(v) + \gamma$. As mentioned before, the best result known for this problem is due to [12, 13] who gave a $(2, 2, 3)$-approximation. In this paper we give a new iterative rounding algorithm which results in a $(2, 2, 2)$-approximation algorithm for Deg-SNDP. We leave open the possibility of a $(2, 2, 1)$-approximation algorithm.

## 3 Main Theorem and the Algorithm

In this section we prove our main theorem.

**Theorem 3.1** (Main Theorem). *There is a polynomial time $(2, 2, 2)$-approximation algorithm for Deg-SNDP.*

The algorithm used appears in Figure 2. We assume that $R$ is weakly super-modular. A *vertex optimal solution* of a LP is an optimal solution which cannot be written as a non-trivial convex combination of two or more feasible solutions to the LP.

It follows from a result of Jain that each iteration of this algorithm can be implemented in polynomial time. It remains to be proved that the algorithm is correct. We first state the main lemma of the paper and then show how it implies the main theorem. The proof of the main lemma is the technical core of the paper and appears in Section 4.

**Lemma 3.2** (Main Lemma). *Given an instance $\mathcal{I} = (G(V, E), W, c, R, b)$, where $R \in \mathbb{Z}_{\geq 0}$ is weakly-super-modular, $b \in \mathbb{Z}_{\geq 0} \cup \{\mathbb{Z} + 1/2\}_{\geq 0}$, let $(x_e)_{e \in E}$ be a vertex optimal solution to the LP of Figure 1 for this instance. Let $E_{>0}$ denote those edges with $x_e > 0$. Then one of the following holds:*

- *there is an edge $e^*$ such that $x_{e^*} = 0$,*
- *or there is an edge $e^*$ such that $x_{e^*} = 1$,*
- *or there is an edge $e^* = \{u, v\} \in E$ such that $1 > x_{e^*} \geq 1/2$, and if $u \in W$, then $b(u) > 1$, and if $v \in W$, then $b(v) > 1$,*
- *or there is a vertex $v \in W$ such that $\deg_{E_{>0}}(v) \leq 2b(v) + 2$.*

Now we see how this lemma implies Theorem 3.1.

*Proof. (of Main Theorem)* Lemma 3.2 implies that each iteration of the algorithm is successful. Further, note that the set of edges $F$ satisfies the requirement function $R$ of $\mathcal{I}$ is true. Hence, when the algorithm ends, since it only picked edges with value at least $1/2$, by a standard argument, the cost of the solution produced by the algorithm is at most $2 \cdot \text{lp}(\mathcal{I}) \leq 2 \cdot \text{opt}(\mathcal{I})$. Hence, the only thing left to prove is that for every $v \in W_0$, its degree in the final integral solution produced by the algorithm is at most $2b(v) + 2$.

1. Given an instance $\mathcal{I} = (G(V, E), W_0, c, R, b)$, initialize

    (a) $F := \emptyset$, $W' := W_0$, $E' := E$, $R'(S) := R(S)$ for all $S \subseteq V$, $b'(v) := b(v)$ for all $v \in W'$.

2. While $F$ is not a feasible solution for $G$ satisfying the connectivity function $R$ do

    (a) Compute a vertex optimal solution $(x_e)_{e \in E}$ to the LP of Figure 1 for the instance $(G(V, E'), W', c, R', b')$. Let $H := \{e \in E' : x_e > 0\}$.

    (b) For every edge $e$ with $x_e = 0$, let $E' := E' \setminus \{e\}$.

    (c) For every edge $e = \{u, v\}$ with $x_e = 1$, let

        i. $F := F \cup \{e\}$,
        ii. If $u \in W'$,
            A. if $b'(u) \neq 3/2$ then $b'(u) := b'(u) - 1$,
            B. else if $b'(u) = 3/2$ then $b'(u) := 1$,
        iii. If $v \in W'$,
            A. if $b'(v) \neq 3/2$ then $b'(v) := b'(v) - 1$,
            B. else if $b'(v) = 3/2$ then $b'(v) := 1$,

    (d) For every edge $e = \{u, v\}$ such that $1/2 \leq x_e < 1$ and if $u \in W'$, then $b'(u) > 1$, and if $v \in W'$, then $b'(v) > 1$, let

        i. $F := F \cup \{e\}$,
        ii. If $u \in W'$, $b'(u) := b'(u) - 1/2$,
        iii. If $v \in W'$, $b'(v) := b'(v) - 1/2$.

    (e) For every degree constrained vertex $v \in W'$ such that $\deg_H(v) \leq 2b'(v) + 2$, let $W' := W' \setminus \{v\}$.

    (f) For every $S \subseteq V$, let $R'(S) := R'(S) - |\delta_H(S)|$.

Figure 2: An Iterative Rounding based algorithm for Deg-SNDP

Consider $v \in W_0$ with degree bound $b(v)$. Suppose that in all the iterations of the algorithm, we picked $n_1$ edges adjacent to $v$ with value 1, and $n_{1/2}$ edges with value in $[1/2, 1)$ adjacent to $v$. In case we picked a 1-ede adjacent to $v$ when $b'(v)$ was $3/2$, then we had decreased $b'(v)$ by $1/2$ and not be 1. Hence, we will count this edge in $n_{1/2}$ and not in $n_1$.

There are two cases:

- If at the point when the algorithm terminated $v$ was still a degree constrained vertex, then its degree in the solution produced by the algorithm would be $n_1 + n_{1/2} < 2b(v)$ as we decreased the degree bound by at least $1/2$ every time we picked an edge adjacent to $v$ and $b'(v) > 0$ at the point of termination of the algorithm.

- If $v \in W_0$ was not a degree constrained vertex when the algorithm terminated, then at some iteration the degree constraint on it had been removed. This happened when in the set of edges $H$ in that iteration $\deg_H(v) \leq 2b'(v) + 2$. Since $v$ had a degree constraint in that iteration, by a similar argument as above we also get that $n_1 + n_{1/2} \leq 2(b(v) - b'(v))$. Now, $\deg_F(v) \leq n_1 + n_{1/2} + 2b'(v) + 2 \leq 2b(v) + 2$.

This completes the proof of Theorem 3.1. □

## 4 Proof of Main Lemma

In this section we prove Lemma 3.2. Consider a vertex optimal solution $(x_e)_{e \in E}$ to the LP for an instance $I = (G(V, E), W, c, R, b)$. Let $E_{>0} \stackrel{\text{def}}{=} \{e : x_e > 0\}$. To prove the lemma we will assume on the contrary all of the following:

1. $0 < x_e < 1$ for each $e \in E_{>0}$.

2. If there is an edge $e = \{u, v\}$ such that $1 > x_e \geq 1/2$, and if $u \in W$ then $b(u) \leq 1$, or if $v \in W$ then $b(v) \leq 1$.

3. For every $v \in W$, $\deg_{E_{>0}}(v) \geq (2b(v) + 2) + 1 \geq 5$.

The proof of this lemma starts with a well known characterization of the vertex optimal solution. The following is standard notation in this setting. A family of sets $\mathcal{L} \subseteq 2^{[V]}$ is said to be *laminar* if for any two sets $S, T \in \mathcal{L}$, either one of them is contained in the other or they are disjoint. In a laminar family, a set $S$ is said to be *child* of $T$ if $T$ is the smallest set containing $S$. ($T$ is called the *parent* of $S$.) Thus, a laminar family can be represented by a forest where the nodes are sets and there is an edge between two sets $S$ and $T$ if one is the child of the other. Let $C(S)$ denote the set of children of $S$. The maximal elements of the laminar family are referred to as the *roots*. The following lemma shows how to derive a laminar family of sets from a vertex optimal solution to the LP of Figure 1.

**Lemma 4.1.** *Given an instance $I = (G(V, E), W, c, R, b)$, where $R$ is weakly-super-modular, $b \in \mathbb{Z}_{\geq 0} \cup \{\mathbb{Z} + 1/2\}_{\geq 0}$, let $(x_e)_{e \in E}$ be a vertex optimal solution to the LP of Figure 1 for this instance. Then, there exists a laminar family of sets $\mathcal{L}$ which partitions into $\mathcal{S}$ and $\mathcal{V}$ such that*

1. *For every $v \in \mathcal{V} \subseteq W$, $x(\delta_E(v)) = b(v) \geq 1/2$ and every $S \in \mathcal{S}$, $x(\delta_E(S)) = R(S) \geq 1$.*

2. $|\mathcal{L}| = |E_{>0}|$.

3. *The vectors $\chi_S$, for $S \in \mathcal{L}$, are linearly independent over the reals.*

*Proof.* The proof follows from the uncrossing method, see [12] and [13]. □

**Notation.** Before we continue with the proof of Lemma 3.2, we need some notation. Let $\mathcal{L}$ be the laminar family associated to the vertex solution $(x_e)_{e \in E}$ as promised by Lemma 4.1. We will refer to a member of $\mathcal{S}$ as a *set* and to a member of $\mathcal{V}$ as a *tight vertex*. An edge $e$ is said to be *heavy* is $1 > x_e \geq 1/2$ and *light* if $0 < x_e < 1/2$. We define the *corequirement* (coreq) of an edge $e$ as $1/2 - x_e$ if $e$ is light and $1 - x_e$ if it is heavy. For a set S, $\text{coreq}(S) = \sum_{e \in \delta(S)} \text{coreq}(e)$. We will say that a set $S$ is *odd* if $\text{coreq}(S) = a + 1/2$, where $a \in \mathbb{Z}_{\geq 0}$, and that it is *even* if $\text{coreq}(S) \in \mathbb{Z}_{\geq 0}$.

We say that a set *owns* an endpoint $u$ of an edge $e = \{u, v\}$ if $S$ is the smallest set in $\mathcal{L}$ containing $u$. For a set $S$ let $c(S)$ denote the number of children of $S$, $l(S)$ the number of endpoints of light edges owned by it, $h(S)$ the number of endpoints of heavy edges owned by it, $l'(S)$ the number of light edges in $\delta(S)$ and $h'(S)$ denote the number of heavy edges in $\delta(S)$. Note that $l'(S) + h'(S) = |\delta(S)|$.

We say that an edge $e$ is incident on a set $S$ if $e \in \delta(S)$. The degree of a set $S$ is defined as the number of edges incident on it, i.e., $\text{degree}(S) \stackrel{\text{def}}{=} |\delta(S)|$. The following fact is easy to see now.

**Fact 4.2.** *A set $S$ has semi-integral corequirement only if $l'(S)$ is odd.*

*Proof.* $\text{coreq}(S) = \sum_{e \in \delta(S)} \text{coreq}(e) = \sum_{e \in \delta(S) \text{ and } e \text{ is light}}(1/2 - x_e) + \sum_{e \in \delta(S) \text{ and } e \text{ is heavy}}(1 - x_e) = l'(S)/2 + h'(S) - f(S)$. All the three terms $l'(S), h'(S), f(S) \in \mathbb{Z}_{\geq 0}$. Therefore, $S$ is semi-integral only if $l'(S)$ is odd. □

*Proof. (of Main Lemma)* We will prove Lemma 3.2 using a counting argument. Initially, we will assign two tokens to each edge. We will redistribute the tokens in such a manner that each member of $\mathcal{L}$ gets at least 2 tokens while the roots get at least 3 tokens. This will give us a contradiction to the fact that $|\mathcal{L}| = |E_{>0}|$ of Lemma 4.1.

**Token distribution scheme.** Initially, we will assign two tokens to each edge. If $e = \{u, v\}$ is a light edge then one of the two tokens assigned to $e$ goes to the smallest set containing $u$ and the other to the smallest set containing $v$. If $e$ is a heavy edge, then w.l.o.g. assume that $u \in W, b(u) = 1$. In this case assign both tokens of $e$ to the smallest set containing $v$.

**Claim 4.3.** *[Token Redsitribution] Consider a tree $\mathcal{T}$ in $\mathcal{L}$ rooted at $S$. The tokens owned by $\mathcal{T}$ can be redistributed in such a way that $S$ gets at least 3 tokens, and each of its descendants gets at least 2 tokens. Moreover, if $\text{coreq}(S) \neq 1/2$, then $S$ gets at least 4 tokens.*

*Proof.* We will prove this claim by induction on the height of $\mathcal{T}$. We start with the base case.

1. If $S$ is a leaf set and it has no heavy edges incident on it then, since $f(S) = \sum_{e \in \delta(S)} x_e \geq 1$ and $\forall e \in \delta(S) : x_e < 1/2$, therefore $S$ has at least 3 edges incident on it and, hence, will collect at least 3 tokens. It will collect exactly 3 tokens when its degree is 3 in which case $\text{coreq}(S) = \sum_{e \in \delta(S)}(1/2 - x_e) = 3/2 - \sum_{e \in \delta(S)} x_e = 3/2 - f(S)$; since $f(S) \in \mathbb{Z}^+$ and $\text{coreq}(S) > 0$ therefore $\text{coreq}(S) = 1/2$. Hence, by the inductive hypothesis it suffices for S to collect only 3 tokens.

2. In the case when $S$ has a heavy edge, say $e_1$, incident on it, it will still have at least 1 other edge incident on it as $f(S) = \sum_{e \in \delta(S)} x_e$ is a positive integer and $\forall e \in \delta(S), x_e < 1$. Recall that, by our assumption a heavy edge must have a tight vertex with degree bound equal to 1 as one of its end points. Since S is a leaf set, the tight vertex cannot be contained in $S$. Therefore, the tight vertex must be that end point of the heavy edge which is not in $S$. By our token distribution scheme, $S$ would get both the tokens from the heavy edge and at least one token from the other edge. Hence, it will get at least 3 tokens. $S$ will get exactly 3 only when it has only one other light edge, say $e_2$, incident on it. In such a case $\text{coreq}(S) = \sum_{e \in \delta(S)} \text{coreq}(e) = (1 - x_{e_1}) + (1/2 - x_{e_2}) = 3/2 - f(S)$; since $f(S) \in \mathbb{Z}_{>0}$ and $\text{coreq}(S) > 0$, therefore $\text{coreq}(S) = 1/2$. Hence, by the inductive hypothesis it suffices for S to collect only 3 tokens.

3. In the case of a tight vertex $v$, by our assumption it has degree 5 and, hence, will collect at least 5 tokens unless it has a heavy edge incident on it, in which case it might have to give both tokens to the smallest set containing the other endpoint but will still be able to collect at least 4 tokens.

This proves the base case. Now, we move on to the general case. Let us consider the case when $S$ is not a leaf in $\mathcal{L}$. If a set has collected $t$ tokens, we will say that it has a *surplus* of $t - 2$. There are four cases:

1. If $S$ has 4 children (either sets or vertices), then $S$ can collect 1 token from each of its children, as from the inductive hypothesis each of its children has a surplus of at least 1. Thus, S can collect 4 tokens for itself.

2. If $S$ has 3 children (either sets or vertices) and if at least one of them has surplus 2, then $S$ can collect 4 tokens for itself. If $S$ owns any end points then again it can collect at least 4 tokens: 1 from each its children and at least 1 from the end point(s) it owns. If all children have a surplus of exactly 1 then $S$ is still able to collect at least 3 tokens and moreover, by the induction hypothesis, all the children of $S$ must have a corequirement of $1/2$. Furthermore, if $S$ owns no endpoints then, using Claim 4.4, we get that $S$ also has a corequirement of $1/2$. Hence, by the induction hypothesis it suffices for $S$ to collect 3 tokens only.

3. If $S$ has 2 children (either sets or vertices) and both of the children have surplus at least 2 then $S$ can collect 4 tokens from them. If one of them, say $S_1$ has surplus exactly 1, then by the induction hypothesis $\mathsf{coreq}(S_1) = 1/2$. In such a case, by using Claim 4.6, it must own at least 1 end point and, hence, can collect at least 3 tokens. It will collect exactly 3 tokens when both the children have a surplus of exactly one (and, hence, both have a corequirement of $1/2$) and $S$ owns exactly one end point (in which case $l(S) + 2h(S) = 1$). In such a case, by Claim 4.4, it suffices for $S$ to collect 3 tokens only.

4. If $S$ has exactly one child, two cases arise:

   – If the child is a set then, by Claim 4.5, $S$ must own at least 2 end points and, hence, it can collect at least 3 tokens: at least one from the surplus of the child and at least 2 from end points it owns. $S$ will collect exactly 3 tokens if its child has a surplus of exactly 1 (which can happen only when the child has a coreqirement of $1/2$) and $S$ owns exactly 2 end points (in which case $l(S) + 2h(S) = 2$). In such a case, by Claim 4.4, it suffices for $S$ to collect 3 tokens only.

   – If the child is a tight vertex $v$ then again 2 cases arise:

     – $v$ has an integral degree constraint: this case can be handled akin to case when the child is a set.
     – $v$ has semi-integral degree constraint : In this case $b(v) \geq 3/2$ as our algorithm maintains the invariant that $b'(v) \in \mathbb{Z}^+ \cup \{\mathbb{Z}^+ + 1/2\}$. By our assumption degree$(v) \geq 2b(v) + 3 \geq 6$. Hence, $\{v\}$ will be able to collect 6 tokens. By the induction hypothesis it requires only 2 tokens for itself and, hence, can give 4 tokens to $S$.

□

Hence, we have proved Lemma 3.2. □

Now we present the proofs of the claims used in this proof.

**Claim 4.4.** *Let $S \in \mathcal{S}$ and suppose $c(S) + l(S) + 2h(S) = 3$. Furthermore, assume that each child of $S$, if any, has a corequirement of $1/2$. Then $\mathsf{coreq}(S) = 1/2$.*

*Proof.* For a light edge $e$, $\mathsf{coreq}(e) = 1/2 - x_e < 1/2$. For a heavy edge $e$, $\mathsf{coreq}(e) = 1 - x_e \leq 1/2$ as $x_e \geq 1/2$. An argument similar to one used in Exercise 23.3 of [18] can be used to show that $S$ is also odd. Using Fact 4.2, $\mathsf{coreq}(S)$ is semi-integral. Now, $\mathsf{coreq}(S) \leq \sum_{C \in C(S)} \mathsf{coreq}(C) + \sum_e \mathsf{coreq}(e)$ where the second summation is over all those edges whose 1 endpoint is owned by $S$. Every term in the first summation is $1/2$ and every term in the second summation is at most $1/2$ (by definition of corequirement, $\mathsf{coreq}(e) \leq 1/2 \ \forall e \in E$). Note that $\mathsf{coreq}(S)$ is a sum of at most 3 terms each of which is at most $1/2$. Therefore, $\mathsf{coreq}(S) \leq 3/2$. Hence, proving that $\mathsf{coreq}(S) < 3/2$ suffices. We do this next.

A term in the second summation will be exactly equal to $1/2$ if the edge corresponding to it is a heavy edge, i.e. $h(S) = 1$. Hence, from the premises of this claim we get that $c(S) + l(S) = 1$. This means that there can be either 1 light edge or 1 child or $S$ contributing to the summation. Either of them will contribute at most half to the summation and, therefore, $\mathsf{coreq}(S) \leq 1$. Considering the case when every term in the second summation is strictly less than $1/2$, $\mathsf{coreq}(S) < 3/2$ and, hence, $\mathsf{coreq}(S) = 1/2$. Now the second summation cannot be empty as then the set of edges incident on $S$ would be exactly all the edges incident on its children. This would contradict the linear independence of the vectors $\chi(\delta(S)) \cup \{\chi(\delta(C))\}_{C \in C(S)}$. □

**Claim 4.5.** *If a set $S$ has only 1 child which is a set, then $S$ owns at least two end points.*

*Proof.* Let $S_1$ be a set which is the only child of $S$. If $S$ owned no end point, that would contradict the linear independence of $\chi(\delta(S))$ and $\chi(\delta(S_1))$. Therefore, $S$ owns at least one endpoint. If $S$ owned exactly one end point (associated with an edge, say $e$) then $x(\delta(S))$ and $x(\delta(S_1))$ would differ by a $x_e$, a fraction, which would contradict the fact that $S$ and $S_1$ are tight sets having integral connectivity requirements. Hence, $S$ owns at least two endpoints, which proves this claim. □

**Claim 4.6.** *If $S$ has two children (either sets or vertices), one of which has a corequirement of $1/2$, then $S$ must own at least one end point.*

*Proof.* Let $C_1$ and $C_2$ be the children of $S$ with $\mathsf{coreq}(C_1) = 1/2$. If $C_1$ were a tight vertex, say $v$, then $\mathsf{coreq}(v) = 1/2$ is equal to $(|\delta(v)|/2 - b(v))$ if $v$ has no tight edge incident on it and is equal to $(|\delta(v)|/2 + 1/2 - b(v))$ if it has a tight edge incident on it. In either case, $|\delta(v)| \leq 2b(v) + 1$ and, hence, we would have removed the degree constraint from $v$. Therefore, $C_1$ cannot be a tight vertex and, hence, has to be a tight set.

Suppose $S$ does not own any end point. Let $f_1 \stackrel{\text{def}}{=} \sum_{e \in \delta(S) \cap \delta(C_1)} \mathsf{coreq}(e)$, $f_2 \stackrel{\text{def}}{=} \sum_{e \in \delta(C_1) \cap \delta(C_2)} \mathsf{coreq}(e)$ and $f_3 \stackrel{\text{def}}{=} \sum_{e \in \delta(S) \cap \delta(C_2)} \mathsf{coreq}(e)$. By definition, $f_1, f_2, f_3 \geq 0$. Since $\chi_{\delta(S)}$, $\chi_{\delta(C_1)}$ and $\chi_{\delta(C_2)}$ are linearly independent, there has to be at least one edge incident on $C_1$ and $C_2$, i.e., $|\delta(C_1) \cap \delta(C_2)| \geq 1$. Therefore, $f_2 > 0$. $f_1 + f_2 = \mathsf{coreq}(C_1) = 1/2$. Now, since $\mathsf{coreq}(C_1)$ is semi-integral and that $C_1$ is a set, $l(C_1)$ must be odd by Fact 4.2.

Further, $l'(S) = l'(C_1) + l'(C_2) - 2l'(\delta(C_1) \cap \delta(C_2))$. Therefore, if $l'(C_2)$ is odd then $l'(S)$ will be even and if $l'(C_2)$ is even then $l'(S)$ will be odd. $l'(C_2)$ and $l'(S)$ having different parities implies that $S$ and $C_2$ have different corequirements: one of them being integral and one being semi-integral. Now, $\mathsf{coreq}(S) - \mathsf{coreq}(C_2) = (f_1 + f_3) - (f_2 + f_3) = f_1 - f_2$. But $f_1, f_2 \geq 0$, $f_1 + f_2 = 1/2$ and $f_2 > 0$ and, hence, $-1/2 < f_1 - f_2 < 1/2$, which implies that $\mathsf{coreq}(S)$ and $\mathsf{coreq}(C_2)$ (both being half-integral and differing by less than $1/2$) must be equal, which contradicts the fact $S$ and $C_2$ have different corequirements. Hence, it cannot be the case that $S$ owns no end point. □

**Possible improvements?** We observe that the techniques developed in this paper cannot be used to prove a $(2, 2, 1)$-approximation algorithm in a straight-forward manner. Suppose we modified Step (e) of Algorithm 2 to one in which we removed the degree constraint from a vertex $v \in W'$ if $\deg_H(v) \leq 2b'(v) + 1$, then we can suitable modify the analysis to show that the modified algorithm is a $(2, 2, 1)$-approximation algorithm. But we show that we cannot prove a $(2, 2, 1)$-approximation guarantee with our techniques.

Consider a part of the set system as shown in Figure 4. $S, C, G \in \mathcal{S}$ and $f(S) = f(C) = f(G) = 1$, $u, v \in \mathcal{T}$ and $b'(u) = b'(v) = 1$. The lines show the edges, the thick lines are the heavy edges and the thin lines are light edges. By our token distribution, we get that the total number of tokens collected by $S$ and its decendants is 10 (as there are 10 endpoints contained in $S$ and its decendants). So $S, C, G, u, v$ will get 2 tokens each, which is clearly insufficient to prove Claim 4.3.

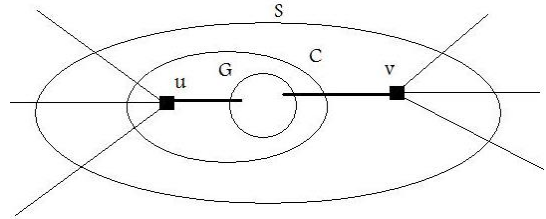

Figure 3: Tight Example

**Extensions.** Our techniques trivially extend to the case when there is a lower bound $l(v)$ on the degree of each vertex $v \in V$. Any degree lower bound constraint can be considered as a connectivity constraint $R(v) = l(v)$ (for the cut $(\{v\}, V \setminus \{v\})$). It can be easily verified that the augmented connectivity function $R$ still remains weakly super-modular. Therefore, any feasible solution to LP of Figure 1 with the augmented $R$ will satisfy all degree lower bounds implicitly.